# Impurity effects on residual zonal flow in deuterium (D)-tritium (T) plasmas

Weixin Guo, Lu Wang,* and Ge Zhuang

State Key Laboratory of Advanced Electromagnetic Engineering and Technology, School of Electrical and Electronic Engineering, Huazhong University of Science and Technology, Wuhan, Hubei 430074, China

*E-mail: luwang@hust.edu.cn

## Abstract

Significant effects of impurities on residual zonal flow (ZF) in deuterium (D)-tritium (T) plasmas are found. When the gyroradius of impurities is larger (smaller) than that of main ions, the intermediate scale (radial wavelength between trapped ion radial width $\rho_{bi}$ and trapped electron radial width $\rho_{be}$) residual ZF level is increased (decreased) due to the presence of various impurities with the tolerance concentration in JET and ITER, even for trace tungsten (W). For short scale (radial wavelength comparable to $\rho_{be}$) region, the residual ZF level is increased by most of the impurities. Moreover, the trend of stronger intermediate residual ZF in D-T plasmas with heavier effective isotope mass is weakened by non-trace impurities, but is not influenced by trace W. These results reveal that the presence of impurities can modify residual ZF, and possibly further affect the ZF regulation of turbulence as well as the associated anomalous transport and confinement in magnetic fusion plasmas. The potential relevance of our findings to experimental observations and simulation results is discussed.

## 1. Introduction.

Zonal flow (ZF) is found to be a ubiquitous phenomenon in both nature and laboratory [1]. It plays an important role in showing Jupiter's zonal winds and understanding the solar dynamo mechanism in solar tachocline. In magnetic confinement plasmas, it is widely recognized that ZF is significant for suppressing micro-turbulence and reducing the anomalous transport. ZF has been observed to play a crucial role in triggering L-H transition, especially at marginal input power [2-5]. Rosenbluth-Hinton (RH) [6] demonstrated that the level of large scale ( $k_r \rho_{bi} \ll 1$, where $k_r$ is the radial wave vector of ZF, $\rho_{bi} = \frac{q}{\sqrt{\epsilon}} \rho_i$ is the trapped ion radial width with $\epsilon$ being the inverse aspect ratio, $q$ being the safety factor and $\rho_i$ being the ion gyroradius) residual ZF driven by ion temperature gradient (ITG) turbulence is not damped by collisionless process, but modified by neoclassical polarization shielding. Its extensions to short wavelength electron-temperature-gradient (ETG) turbulence [7, 8], arbitrary radial wavelengths [9, 10], shaped tokamak geometry, collisional case [11, 12] and stellarators [13, 14] were reported. However, all these previous works did not address the effects of non-hydrogenic ions i.e., *impurities* on residual ZF.

Impurities produced mainly from the plasma facing materials interaction or helium ash in deuterium (D)-tritium (T) plasmas may significantly influence the success of fusion through causing energy loss and fuel dilution. But, on the other hand, tokamak experiments have also widely demonstrated the significant improvement of global confinement by injecting impurities [15]. Physical interpretations about the reason why impurities affect the plasma performance are complicated. One possibility is that impurities influence on ZF, and thus the regulation of micro-turbulence and turbulent transport by ZF may be further affected. Actually, enhancement of flow shearing rate and improvement of confinement with light impurity injection have been observed in DIII-D [16]. Since International Thermonuclear Experimental Reactor (ITER) and some present-day tokamaks such as JET, KSTAR, EAST use tungsten (W) in the divertors, the study of W effects on plasma performance becomes an important topic. The typical tolerance concentration (ratio of impurity density to electron density $f_c = n_{0z}/n_{0e}$ with $n_{0e}$ and $n_{0z}$ being the electron and impurity equilibrium densities) of W is the order of $10^{-4}$ in JET [17], which indicates that W behaves as a trace impurity. Theoretically, it was found that highly charged impurities significantly enhance the collisional damping of residual ZF, but do not affect the collisionless large scale RH residual ZF [18]. However, how the highly charged impurities affect arbitrary scale residual ZF is still lacking. Moreover, effects of the light or medium-mass impurities with finite concentration (called non-trace impurities) on arbitrary scale residual ZF are also remained to be investigated, especially in the plasmas composed mainly by D-T mixtures. This is very important for burning plasmas in ITER and DEMO devices. The present work therefore aims to contribute the knowledge about the effects of various impurities on arbitrary scale residual ZF in D-T plasmas, which may have potential relevance to the effects of impurities on turbulence and confinement in magnetic fusion plasmas.

The energy confinement time $\tau_E$ has been observed to be improved as changing from hydrogen (H) to D or D-T plasmas in the different operation regimes [19]. This is the well-known isotopic effects. While, these experimental results are in contradiction with the prediction of the so-called Gyro-Bohm scaling [15] which shows a hydrogenic isotope mass dependence, i.e., $\tau_E \sim A_i^{0.50}$ with $A_i$ being the isotope mass number of main ions. Although some physical mechanisms were proposed to explain the contradiction [20-22], the fully understanding is still not achieved yet. Gyrokinetic simulation in Ref. [23] revealed that the thermal diffusivity decreases with the increase of hydrogenic isotope mass in ITG turbulence, which is partly attributed to the linear growth rate decreasing with the increase of hydrogenic isotope mass. In the presence of impurities, this favorable isotopic scaling is also reported in both ITG and trapped electron mode (TEM) turbulence [24-26]. The effects of impurities on hydrogenic isotope mass dependence of confinement time in saturated Ohmic confinement (SOC) regime in ASDEX were observed [27]. Recently, both gyrokinetic [28, 29] and gyro-fluid [30] simulations have investigated the hydrogenic isotope mass dependence of ZF and geodesic acoustic mode (GAM). In Ref. [31], a stronger intermediate residual ZF level in D plasmas than that in H plasmas was analytically addressed, and possible relevance to the isotopic effects was also discussed. This analytical result was also verified in gyrokinetic simulations [32]. But all these simulation and analytical works on hydrogenic isotope mass dependence of residual ZF did not take impurities into account. Moreover, the mixing ratio of D and T will be changed step by step for D-T operation in ITER according the newest research plan. More fraction of T is favorable for achieving high confinement mode because of lower power threshold in T plasmas, which may be attributed to higher ZF in T plasmas than that in D plasmas. Both isotopic effects and impurities are very important issues for the second D-T campaign in JET and extrapolation to ITER. Therefore, the other goal of this work is to examine the impurity effects on the hydrogenic isotope mass dependence of residual ZF in plasmas with different D-T mixing ratio. This may be relevant to the impurity effects on the hydrogenic isotope mass dependence of confinement time, and provide a possible clue to the optimization of D-T mixing ratio from the viewpoint of interplay between impurities and ZF.

In this work, we systematically study the effects of impurities on arbitrary scale residual ZF in collisionless D-T plasmas. The arbitrary radial wavelength includes three limiting cases: intermediate scale (radial wavelength between trapped ion radial width $\rho_{bi}$ and trapped electron radial width $\rho_{be}$), short scale (radial wavelength that are comparable to $\rho_{be}$) and large scale (radial wavelengths is much larger than $\rho_{bi(z)}$), where $\rho_{b\alpha} = \frac{q}{\sqrt{\epsilon}}\rho_\alpha$ is the trapped particles' radial width with $\rho_\alpha$ being the gyroradius of species $\alpha$ and $\alpha = e, i, z$ corresponding to electron, ion and impurity. The general expression for residual ZF is derived by including dynamics of three species, i.e., electrons, ions and impurities. We discuss two kinds of impurities. One is the light or medium-mass ($Z \leq 18$, $A_z \leq 40$) impurity with finite concentration according to ITER, particularly including the high temperature helium impurity ($He^{2+}$) from D-T reaction with its tolerance concentration. Here, $Z$ and $A_z$ represent impurity charge number and impurity mass number, respectively. The other is the highly charged trace W, which can be produced from the divertor of ITER and some present-day tokamaks such as JET. Taking impurities into account, we find significant decrease (increase) of intermediate scale residual ZF when $\rho_z$ is smaller (larger) than $\rho_{i,eff}$ with $\rho_{i,eff}$ being the effective ion gyroradius, even to 35% (15%). Surprisingly, we also find that the intermediate scale residual ZF in D-T plasmas is decreased about 15% in the presence of high-$Z$ W even with trace concentration. Moreover, the decreasing (increasing) trend due to the presence of impurities is strengthened with the decrease (increase) of $\frac{\rho_z^2}{\rho_{i,eff}^2} \sim \frac{A_z}{A_{i,eff}Z^2}\frac{T_z}{T_i}$, where $T_i$ and $T_z$ are the ion and impurity temperature, respectively, $A_{i,eff}$ is the effective isotope mass number of D-T plasmas. The decreasing and increasing trends are also strengthened by further increase of $f_c$. Impurity effects on short scale residual ZF are parametric dependence. The increased (decreased) residual ZF due to the presence of impurities possibly leads to lower (stronger) anomalous transport and better (inferior) confinement. It is also found that the trend of stronger intermediate scale residual ZF in heavier D-T plasmas is weakened with the increase of $f_c$, $A_z$ and $Z$ for light and medium-mass non-trace impurities, while the change of $Z$ for trace W has very weak influence on this trend. Then, the potential relevance of our findings to experimental observations and simulation results is discussed.

This paper is organized as follows. In section 2, the general expression for arbitrary radial wavelength residual ZF including impurities is presented. In section 3, the effects of various impurities on residual ZF in D-T plasmas are analyzed in detail. Finally, a summary and some discussions are given in section 4.

## 2. General expression for residual ZF with impurities.

As pointed out in the original RH residual ZF model [6], the initial charge density perturbation is accompanied by a potential perturbation because of quasi-neutrality condition $e\left(\delta n_{i,k} - \delta n_{e,k}\right) = -\rho_k^{NL}(0)$, where $e$ is the elementary charge, $\delta n_{i,k}$ and $\delta n_{e,k}$ are the ion and electron perturbed density, respectively, and $\rho_k^{NL}(0)$ is the initial nonlinear charge source. The initial zonal potential perturbation $\phi_{ZF,k}(t=0)$ is built by the classical polarization shielding (leading to the particle departure from the gyrocenter) within a time scale of several ion gyroperiods, i.e.,

$$e\left[\left(1 - \frac{1}{n_{0i}}\left\langle\int d^3 v F_{0i} \overline{J_{0i}^2}\right\rangle\right)\frac{en_{0i}}{T_i} + \left(1 - \frac{1}{n_{0e}}\left\langle\int d^3 v F_{0e} \overline{J_{0e}^2}\right\rangle\right)\frac{en_{0e}}{T_e}\right]\delta\phi_{ZF,k}(t=0) = \rho_k^{NL}(0). \quad (1)$$

Here, $n_{0i}$ is the ion equilibrium density, $T_e$ is the electron temperature, $F_{0i(e)}$ is the equilibrium distribution function of ions (electrons), $J_{0i(e)}$ is the zeroth-order Bessel function for ions (electrons). The meaning of other symbols have been explained in previous section. A few of bounce periods later, the neoclassical polarization shielding originating from the gyrocenter departure from bounce center modifies the initial zonal potential perturbation. The long-time behavior of zonal potential perturbation $\phi_{ZF,k}(t=\infty)$ is then determined by the summation of classical polarization and neoclassical polarization

$$e\left[\left(1-\frac{1}{n_{0i}}\left\langle\int d^3v F_{0i}J_{0i}e^{-iQ_i}\overline{e^{iQ_i}J_{0i}}\right\rangle\right)\frac{en_{0i}}{T_i}+\left(1-\frac{1}{n_{0e}}\left\langle\int d^3v F_{0e}J_{0e}e^{-iQ_e}\overline{e^{iQ_e}J_{0e}}\right\rangle\right)\frac{en_{0e}}{T_e}\right]$$
$$\times\delta\phi_{ZF,k}(t=\infty)=\rho_k^{NL}(0). \tag{2}$$

Here, $\langle...\rangle$ represents the flux surface average, $Q_{i(e)}=\frac{v_{\|,i(e)}IS'}{\Omega_{i(e)}}$ with $v_{\|,i(e)}$ being the ion (electron) parallel velocity, and $I=RB_\varphi$ with $R$ being the major radius and $B_\varphi$ being the toroidal magnetic field, $S$ is the eikonal under assuming all the perturbed quantities to be in an eikonal form $\delta\phi=\sum_k\delta\phi_k e^{iS}$ [6] and $S'$ represents the gradient of $S$, $\Omega_{i(e)}=\frac{eB}{m_{i(e)}c}$ is the ion (electron) cyclotron frequency with $B$ being the total magnetic field, $m_{i(e)}$ being the ion (electron) mass and $c$ being the light velocity. Then, the residual ZF level $R_{ZF}$ defined as the ratio of $\phi_{ZF,k}(t=\infty)$ to $\phi_{ZF,k}(t=0)$, which is a dimensionless quantity, is then written as

$$R_{ZF}=\frac{\tau_i\frac{n_{0i}}{n_{0e}}\left(1-\frac{1}{n_{0i}}\left\langle\int d^3v F_{0i}\overline{J_{0i}^2}\right\rangle\right)+\left(1-\frac{1}{n_{0e}}\left\langle\int d^3v F_{0e}\overline{J_{0e}^2}\right\rangle\right)}{\tau_i\frac{n_{0i}}{n_{0e}}\left(1-\frac{1}{n_{0i}}\left\langle\int d^3v F_{0i}J_{0i}e^{-iQ_i}\overline{e^{iQ_i}J_{0i}}\right\rangle\right)+\left(1-\frac{1}{n_{0e}}\left\langle\int d^3v F_{0e}J_{0e}e^{-iQ_e}\overline{e^{iQ_e}J_{0e}}\right\rangle\right)}, \tag{3}$$

where $\tau_i=T_e/T_i$. More detailed interpretations can be also found in Ref. [11]. Previous works either focus on ion polarization shielding [6] or include the polarization shielding of electrons [7-10] as well. But, for plasmas containing impurities, the contribution from impurities to the polarization shielding should be also included. Therefore, the general expression for residual ZF with impurities is given by

$$R_{ZF} = \frac{\tau_i \frac{n_{0i}}{n_{0e}}\left(1-\frac{1}{n_{0i}}\left\langle\int d^3v F_{0i}\overline{J_{0i}^2}\right\rangle\right)+Z^2\tau_z\frac{n_{0z}}{n_{0e}}\left(1-\frac{1}{n_{0z}}\left\langle\int d^3v F_{0z}\overline{J_{0z}^2}\right\rangle\right)+\left(1-\frac{1}{n_{0e}}\left\langle\int d^3v F_{0e}\overline{J_{0e}^2}\right\rangle\right)}{\left[\tau_i\frac{n_{0i}}{n_{0e}}\left(1-\frac{1}{n_{0i}}\left\langle\int d^3v F_{0i}J_{0i}e^{-iQ_i}\overline{e^{iQ_i}J_{0i}}\right\rangle\right)+Z^2\tau_z\frac{n_{0z}}{n_{0e}}\left(1-\frac{1}{n_{0z}}\left\langle\int d^3v F_{0z}J_{0z}e^{-iQ_z}\overline{e^{iQ_z}J_{0z}}\right\rangle\right)\right] + \left(1-\frac{1}{n_{0e}}\left\langle\int d^3v F_{0e}J_{0e}e^{-iQ_e}\overline{e^{iQ_e}J_{0e}}\right\rangle\right)}$$

$$= \frac{\tau_i\frac{n_{0i}}{n_{0e}}\chi_{i,cl}+Z^2\tau_z\frac{n_{0z}}{n_{0e}}\chi_{z,cl}+\chi_{e,cl}}{\tau_i\frac{n_{0i}}{n_{0e}}\left(\chi_{i,cl}+\chi_{i,nc}\right)+Z^2\tau_z\frac{n_{0z}}{n_{0e}}\left(\chi_{z,cl}+\chi_{z,nc}\right)+\chi_{e,cl}+\chi_{e,nc}}. \quad (4)$$

Here, $\chi_{\alpha,cl} = 1 - \frac{1}{n_{0\alpha}}\left\langle\int d^3v F_{0\alpha}\overline{J_{0\alpha}^2}\right\rangle$, $\chi_{\alpha,nc} = \frac{1}{n_{0\alpha}}\left\langle\int d^3v F_{0\alpha}\left(\overline{J_{0\alpha}^2}-J_{0\alpha}e^{-iQ_\alpha}\overline{e^{iQ_\alpha}J_{0\alpha}}\right)\right\rangle$

with $\alpha = e, i, z$ are the classical and neoclassical polarization shieldings, respectively. The neoclassical polarization density is defined as the difference between the gyrocenter density and the bounce center density based on modern gyrokinetic and bounce kinetic theory, and it can be obtained from the pull-back transformation from bounce center to gyrocenter by keeping both the finite Larmor radius (FLR) effects and finite orbit width (FOW) effects. $\tau_z = T_e/T_z$, and the other symbols for impurities have the similar meanings to those for ions and electrons. The equilibrium quasi-neutrality condition indicates $\frac{n_{0i}}{n_{0e}} = 1 - Zf_c$. For simplicity, Eq. (4) can be rewritten as

$$R_{ZF} = \frac{\sum_\alpha g_\alpha \chi_{\alpha,cl}}{\sum_\alpha g_\alpha\left(\chi_{\alpha,cl}+\chi_{\alpha,nc}\right)}. \quad (5)$$

Here, $g_e = 1$, $g_i = \tau_i(1-Zf_c)$, $g_z = \tau_z Z^2 f_c$ are weighting factors corresponding to electrons, ions and impurities.

The classical polarization shielding has the well-known form, i.e., $\chi_{\alpha,cl} = 1 - \Gamma_0\left(k_r^2\rho_\alpha^2\right)$ with $\Gamma_0 = I_0\left(k_r^2\rho_\alpha^2\right)e^{-k_r^2\rho_\alpha^2}$ and $I_0$ being the zeroth-order modified Bessel function. In the high aspect ratio concentric circular geometry, the generalized expression of $\chi_{\alpha,nc}$ for arbitrary scale was constructed by adding the inverse of three asymptotic forms, and then taking the inverse of the summation [10],

$$\chi_{\alpha,nc} = \left\{ \frac{\epsilon^{1/2}}{1.83 q^2 k_r^2 \rho_\alpha^2} + \left[1 + \frac{\sqrt{8\epsilon}}{\pi}\Gamma_{tr}' + \left(1 - \frac{\sqrt{8\epsilon}}{\pi}\right)\Gamma_p'\right]\frac{1}{1+k_r^2\rho_\alpha^2} + \sqrt{\frac{\pi^3}{2}} k_r \rho_\alpha \left[1 + \frac{\sqrt{8\epsilon}}{\pi}\Gamma_{tr} + \left(1 - \frac{\sqrt{8\epsilon}}{\pi}\right)\Gamma_p\right]\frac{k_r^2\rho_\alpha^2}{1+k_r^2\rho_\alpha^2} \right\}^{-1}. \qquad (6)$$

Here, $\sqrt{8\epsilon}/\pi$ and $1-\sqrt{8\epsilon}/\pi$ are the fractions of trapped particles and passing particles, respectively; $\Gamma_{tr} \simeq 0.92/(\sqrt{\pi\epsilon} k_r \rho_{\theta\alpha})$ ( $\Gamma_{tr}' = 2\Gamma_{tr}/\pi$ ) and $\Gamma_p = 1/(2\sqrt{\pi\epsilon} k_r \rho_{\theta\alpha})$ ($\Gamma_p' = 2\Gamma_p/\pi$) are related to the neoclassical polarization for trapped particles and passing particles, respectively, where $\Gamma_{tr}$ and $\Gamma_p$ are inversely proportional to the banana width $\rho_{b\alpha} = \frac{q}{\sqrt{\epsilon}}\rho_\alpha$ and the radial deviation from the flux surface for a strongly passing particle $q\rho_\alpha$, respectively, in the limit of $k_r\rho_{\theta\alpha} > k_r\rho_\alpha \gg 1$ ( $k_r\rho_\alpha \ll 1$ and $k_r\rho_{\theta\alpha} \gg 1$ ); the factor $\sqrt{\pi^3/2}k_r\rho_\alpha$ from the inverse of FLR effects was kept in the limit of $k_r\rho_{\theta\alpha} > k_r\rho_\alpha \gg 1$. In the opposite limit, i.e., $k_r\rho_\alpha < k_r\rho_{\theta\alpha} \ll 1$, Eq. (6) will be reduced to RH neoclassical polarization with a slightly different coefficient. The main point for calculating $\chi_{\alpha,nc}$ is that the orbit width (Larmor radius or banana width or the radial deviation from the flux surface for passing particle) comparable to the wavelength of fluctuations is the most relevant to the polarization shielding. This is why the FLR effects on neoclassical polarization cannot be ignored in the limit of $k_r\rho_{\theta\alpha} > k_r\rho_\alpha \gg 1$ with $\rho_{\theta\alpha} = \frac{q}{\epsilon}\rho_\alpha$ being the poloidal gyroradius. The interested readers can refer to the Ref. [10] for the details of derivation processes. The general neoclassical polarization which was used only for electrons and ions in previous work can be also applicable to impurities.

## 3. Effects of various impurities on residual ZF in D-T plasmas.

In this part, we investigate how various impurities affect the residual ZF in D-T plasmas. We use the following typical parameters: $q=1.4$, $\epsilon=0.2$. The isotopic fueling ratios $f_D = n_{0D}/(n_{0D}+n_{0T})$ and $f_T = n_{0T}/(n_{0D}+n_{0T})$ are 50%+50% in section 3.1, and they are changed from $f_T > f_D$ to $f_T < f_D$ in section 3.2. Especially, the tolerance concentrations for He$^{2+}$ from D-T reaction, Be$^{4+}$, Ar$^{18+}$ in ITER are 10%, 2% and 0.16% [15], respectively, and $f_c = 10^{-4}$ for trace W [17]. Normally, we assume $\tau_z = \tau_i = 1$ $(T_z = T_i = T_e)$ for most of impurities. Interestingly, the fusion products, i.e., energetic alpha particles dominantly heat the electrons [33], and then exchange energy to ions by collisions. Meanwhile, the slowing down time of alpha particles is typically longer than the energy exchange time, so we assume $\tau_i = 1$ $(T_i = T_e)$, $\tau_z \leq 1$ $(T_e \leq T_z)$ for high temperature He$^{2+}$ from D-T reaction.

## 3.1 Residual ZF in 50%+50% D-T plasmas with various impurities

In this subsection, we present the effects of various impurities on residual ZF in 50%+50% D-T plasmas, i.e., $A_{i,eff} = 2.5$. From Eq. (5), it can be seen that the impurities affect residual ZF mainly through their weighting factors and polarization shieldings. In figure 1(a), we compare the classical and neoclassical polarization shieldings of high temperature He$^{2+}$ from D-T reaction with $\tau_z = 0.1$ (blue lines) and $\tau_z = 1$ (green lines) with those of electrons (black lines) and ions (red lines). Both $\chi_{z,cl}$ and $\chi_{z,nc}$ are comparable to the corresponding components of ions. When $\rho_z > (<) \rho_{i,eff}$, $\chi_{z,nc}$ changes faster (slower) than $\chi_{i,nc}$. Figure 1(b) shows the effects of light non-trace Be$^{4+}$, medium-mass Ar$^{18+}$ with concentration in ITER and W with trace concentration in JET on residual ZF in D-T plasmas, respectively. It is obvious that the levels of intermediate and short scale residual ZFs are affected by impurities even for high-Z W with the trace concentration. To clearly illustrate the different residual ZF levels between the cases with and without impurities, we discuss how the variations of $A_z$, $Z$, $f_c$ and $\tau_z$ affect the ratio $R_{ZF,z} / R_{ZF,0}$ in detail in the following, where $R_{ZF,z}$ and $R_{ZF,0}$ are residual ZF levels with and without impurities, respectively. The main results are shown in figures 2, 3, 4 and 5.

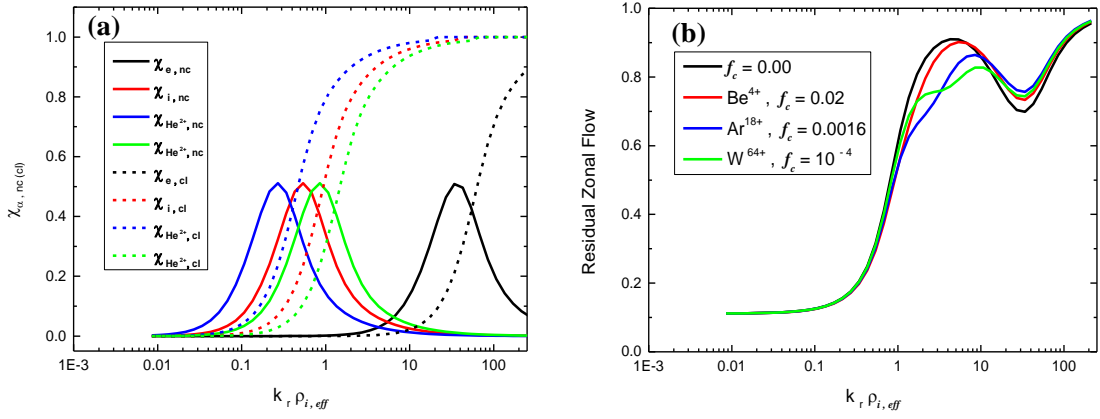

Figure 1. (a) The classical (dashed lines) and neoclassical (solid lines) polarization shieldings for electrons (black lines), ions (red lines) and high temperature $He^{2+}$ with $\tau_z = 0.1$ (blue lines) and $\tau_z = 1$ (green lines) and (b) the residual ZF level without impurities (black line) and with non-trace $Be^{4+}$ (red line), $Ar^{18+}$ (bule line) and trace $W^{64+}$ (green line) as a function of $k_r\rho_{i,eff}$. The impurity concentrations for $Be^{4+}$, $Ar^{18+}$ and $W^{64+}$ are $0.02$, $0.0016$ and $10^{-4}$, respectively.

Intermediate scale residual ZF, which may be driven by TEM turbulence [34-37], is decreased by the presence of impurities with $\rho_z < \rho_{i,eff}$ as shown in figures 2 and 3. The residual ZF can be even decreased about 30% for $Ne^{10+}$ with $f_c = 0.01$. Because the contribution from electron is sub-dominant for the intermediate scale residual ZF, the ratio between residual ZF levels with and without impurities then reduces to

$$\frac{R_{ZF,z}}{R_{ZF,0}} \approx \frac{1+\frac{\chi_{i,nc}}{\chi_{i,cl}}}{1+\frac{g_i\chi_{i,nc}+g_z\chi_{z,nc}}{g_i\chi_{i,cl}+g_z\chi_{z,cl}}}$$

$$= \frac{\frac{\chi_{i,cl}}{\chi_{i,nc}}+1}{\frac{\chi_{i,cl}}{\chi_{i,nc}}+\frac{g_i+g_z\frac{\chi_{z,nc}}{\chi_{i,nc}}}{g_i+g_z\frac{\chi_{z,cl}}{\chi_{i,cl}}}}. \quad (7)$$

The influence from $\chi_{z,cl} < \chi_{i,cl}$ for $\rho_z < \rho_{i,eff}$ is stronger than that from the relationship between $\chi_{z,nc}$ and $\chi_{i,nc}$ in this region. This leads to lower level of residual ZF in the presence of impurities with $\rho_z < \rho_{i,eff}$ as can be seen from Eq. (7). The lower trend is strengthened by the decrease of the ratio $\frac{\rho_z^2}{\rho_{i,eff}^2} \sim \frac{A_z}{A_{i,eff} Z^2} \frac{T_z}{T_i}$ and the increase of $f_c$ as shown in figure 2 (a) and (b). In figure 3, the intermediate scale residual ZF level is decreased about 20% for Ar$^{18+}$ with $f_c = 0.16\%$. But the decreasing trend is not affected by the variation of $T_e/T_{i,z}$ with $T_z = T_i$, because $\frac{\rho_z^2}{\rho_{i,eff}^2}$ does not change with $T_e/T_{i,z}$. These results reveal that impurities have significant influences on intermediate residual ZF. It is believed that ZF can suppress micro-turbulence and reduce the anomalous transport. Therefore, it is possible that impurities may further affect ZF regulation of turbulence and transport. According to the experimental scaling laws $\tau_E \propto Z_{eff}^{-0.27} A_i^{0.50}$ for SOC plasmas [27] with $Z_{eff}$ being the effective charge number, higher $Z_{eff}$ could result in worse confinement. In our results, impurities with higher $f_c$ and higher Z (corresponding to higher $Z_{eff}$) lead to lower level of residual ZF in the intermediate region. The lower level of residual ZF may be relevant to worse confinement, which is consistent with the indications of experimental scaling laws [27].

In short wavelength regime, residual ZF is increased by the presence of impurities as shown in figures 2 and 3. In this region, the residual ZF may be driven by ETG turbulence, where electron polarization shielding should be taken into account. Both impurities and ions can be assumed to be adiabatic, i.e., $\chi_{z,cl}$ and $\chi_{i,cl}$ approach to unity, while $\chi_{z,nc}$ and $\chi_{i,nc}$ decrease to zero. Then, the ratio of residual ZF between the cases with and without impurities can be reduced as

$$\frac{R_{ZF,z}}{R_{ZF,0}} = \frac{1 + \frac{\chi_{e,nc}}{\tau_i + \chi_{e,cl}}}{1 + \frac{\chi_{e,nc}}{g_i + g_z + \chi_{e,cl}}}. \tag{8}$$

Therefore, impurity effects on short scale residual ZF are mainly through the summation of ion and impurity weighting factors, $g_i + g_z = \tau_i + Zf_c(\tau_z Z - \tau_i)$, which reduces to $\tau_i$ in the absence of impurities. For fully ionized impurities with $\tau_i = \tau_z$, the summation is greater than $\tau_i$, i.e., $g_i + g_z - \tau_i = Zf_c\tau_z(Z-1) > 0$, leading to $\frac{R_{ZF,z}}{R_{ZF,0}} > 1$ as can be seen from Eq. (8).

Higher $Z$, $f_c$ and $\tau_z$ correspond to greater summations, and hence result in higher levels of residual ZF. Although impurity effects on residual ZF in this short wavelength region are relatively weaker as compared to intermediate scale, the increase of residual ZF caused by impurities may lead to better energy confinement. Especially, this may be important for burning plasmas such as ITER, where energetic alpha particles dominantly heat electrons and ETG turbulence is a plausible candidate channel for electron transport.

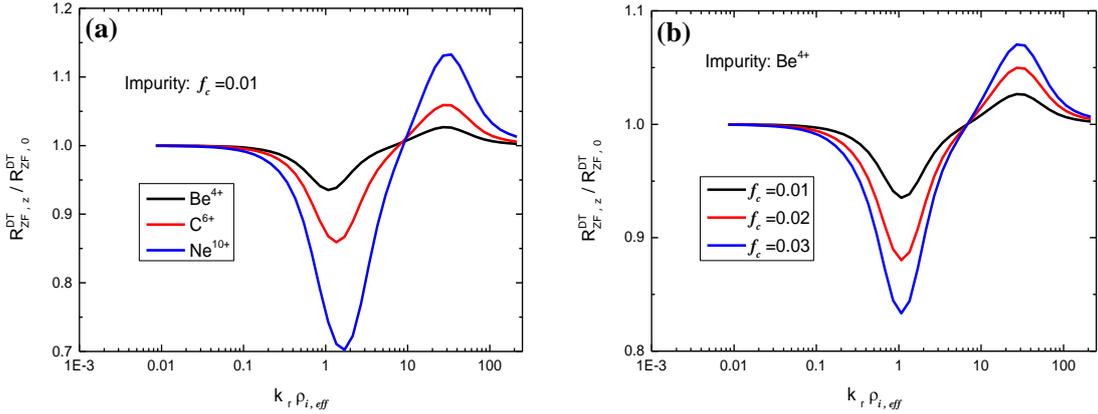

Figure 2. Ratio between $R_{ZF}$ for plasmas with and without impurities as a function of $k_r\rho_{i,eff}$ for different fully ionized impurities with $f_c=0.01$ in (a) and different $f_c$ with $Be^{4+}$.

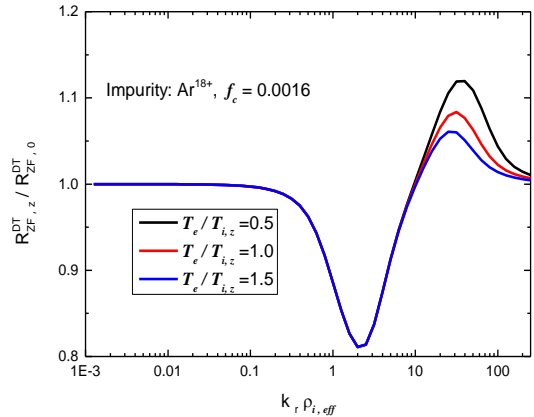

Figure 3. Ratio between $R_{ZF}$ for plasmas with and without impurities as a function of $k_r \rho_{i,eff}$ for different $T_e/T_{i,z}$. The impurity is chosen as $Ar^{18+}$ with $T_z = T_i$ and $f_c = 0.0016$.

Interestingly, figure 4 shows that the intermediate scale residual ZF is increased (decreased) about 12% (35%) by high temperature $He^{2+}$ from D-T reaction with $f_c = 0.1$ when $\frac{\rho_z^2}{\rho_{i,eff}^2} \sim \frac{A_z}{A_{i,eff} Z^2} \frac{T_z}{T_i} > (<) 1$. It seems that the relationship between $\rho_z$ and $\rho_{i,eff}$ plays a key role in determining whether the impurities increase or decrease the intermediate scale residual ZF. Here, $\tau_i = 1$ ($T_i = T_e$) and $\tau_z \leq 1$ are assumed for high temperature $He^{2+}$ from 50%+50% D-T reaction. This might be a good news for suppressing TEM turbulence with higher temperature $He^{2+}$ ($\tau_z < 0.4$) in burning plasmas such as ITER. However, in short wavelength regime, the summation of weighting factors is smaller (larger) than unity for $\tau_z < (>) 0.5$, leading to $\frac{R_{ZF,z}}{R_{ZF,0}} < (>) 1$ as shown in figure 4. The higher temperature $He^{2+}$ ($\tau_z < 0.5$) may be unfavorable for good confinement in the short scale region.

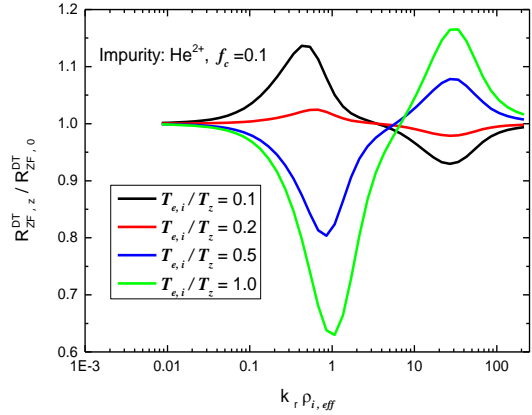

Figure 4. Ratio between $R_{ZF}$ for plasmas with and without high temperature $He^{2+}$ from D-T reaction as a function of $k_r \rho_{i,eff}$ for different $T_{e,i}/T_z$. $T_e = T_i$ and $f_c = 0.1$ are assumed.

Surprisingly, the intermediate scale residual ZF is decreased about 15% even for trace $W^{64+}$ as shown in figure 5. This is because $\frac{\rho_z^2}{\rho_{i,eff}^2} \approx 0.018$ is much smaller than unity for trace $W^{64+}$.

The trace W can enhance the short scale residual ZF, which is similar to the results of non-trace impurities in figures 2 and 3. The non-ignorable effects of W in these two regions are mainly because the impurity weighting factor $g_z = Z^2 f_c \approx 0.41$ for $\tau_z = 1$ is considerable, although $f_c$ is very small. The influences of W on residual ZF may be important for investigating the effects of W on energy confinement in the second D-T campaign in JET and ITER.

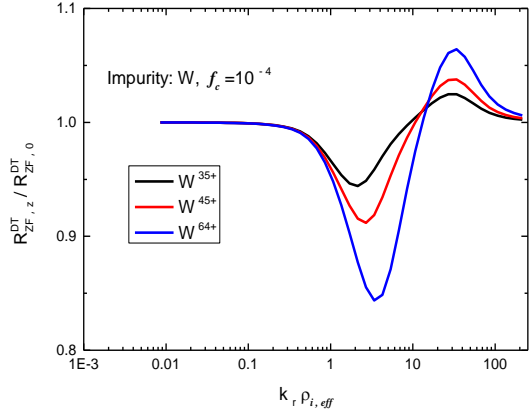

Figure 5. Ratio between $R_{ZF}$ for plasmas with and without trace W as a function of $k_r \rho_{i,eff}$ for different Z. The concentration of W is $10^{-4}$.

In large scale region corresponding to $k_r \rho_{bi(z)} \ll 1$, figures 2, 3, 4 and 5 all show that the introduction of impurities has very weak influence on the residual ZF level. For long wavelength residual ZF, which may be driven by ITG turbulence, electrons can be assumed to be adiabatic. The classical and neoclassical polarization shielding of ions and impurities in this long wavelength limit can be written as $\chi_{i(z),cl} = k_r^2 \rho_{i,eff(z)}^2$, and $\chi_{i(z),nc} = 1.83 q^2 k_r^2 \rho_{i,eff(z)}^2 / \epsilon^{1/2}$, leading to $\frac{\chi_{z,nc}}{\chi_{i,nc}} \approx \frac{\chi_{z,cl}}{\chi_{i,cl}}$. Therefore, the ratio $\frac{R_{ZF,z}}{R_{ZF,0}}$ approaches to unity as can be seen from Eq. (7). In other words, the impurity effects on large scale residual ZF are very weak. This is consistent with Ref. [18] for highly charged impurity. Here, we further point out that the effects of the light non-trace impurities on the large scale residual ZF are also invisible.

## 3.2 Impurity effects on the effective isotope mass dependence of

**residual ZF in D-T plasmas.**

In this subsection, we vary the mixing ratios $f_D$ and $f_T$ to study impurity effects on the $A_{i,eff}$ dependence of residual ZF. $R_{ZF,z}^{0.1D+0.9T}$, $R_{ZF,z}^{0.3D+0.7T}$, $R_{ZF,z}^{0.5D+0.5T}$, $R_{ZF,z}^{0.7D+0.3T}$, $R_{ZF,z}^{0.9D+0.1T}$ separately represent the residual ZF levels corresponding to plasmas with different mixing ratios in the presence of impurities. Figures 6 and 7 show $\frac{R_{ZF,z}^{0.1D+0.9T}}{R_{ZF,z}^{0.5D+0.5T}} > \frac{R_{ZF,z}^{0.3D+0.7T}}{R_{ZF,z}^{0.5D+0.5T}} > 1$ and $\frac{R_{ZF,z}^{0.9D+0.1T}}{R_{ZF,z}^{0.5D+0.5T}} < \frac{R_{ZF,z}^{0.7D+0.3T}}{R_{ZF,z}^{0.5D+0.5T}} < 1$, i.e., the larger $A_{i,eff}$ corresponds to stronger residual ZF in the presence of various impurities in the intermediate scale region. The $A_{i,eff}$ dependence of residual ZF with impurities is similar to the case in the absence of impurities [31]. As discussed in previous subsection, the electron contribution to residual ZF in the intermediate region is sub-dominant, therefore, the residual ZF can be given by

$$R_{ZF,z} \approx \frac{1}{1 + \frac{g_i \chi_{i,nc} + g_z \chi_{z,nc}}{g_i \chi_{i,cl} + g_z \chi_{z,cl}}}. \tag{9}$$

From Eq. (9), it can be seen that the $A_{i,eff}$ dependence of residual ZF is mainly because the influence from larger $\chi_{f_D D + f_T T, cl}$ for larger $A_{i,eff}$ is stronger than that from the corresponding relationship between $\chi_{f_D D + f_T T, nc}$ [31]. To further present the impurity effects on the $A_{i,eff}$ dependence of residual ZF, we focus on how impurities with different $A_z$, $Z$ and $f_c$ affect the above ratios in the intermediate wavelength region.

In figure 6, both $\frac{R_{ZF,z}^{0.1D+0.9T}}{R_{ZF,z}^{0.5D+0.5T}}$ and $\frac{R_{ZF,z}^{0.3D+0.7T}}{R_{ZF,z}^{0.5D+0.5T}}$ ($\frac{R_{ZF,z}^{0.7D+0.3T}}{R_{ZF,z}^{0.5D+0.5T}}$ and $\frac{R_{ZF,z}^{0.9D+0.1T}}{R_{ZF,z}^{0.5D+0.5T}}$) are slightly reduced (enhanced) by the presence of non-trace impurities as compared to the cases without impurities (the lines with $f_c = 0.00$). In other words, the trend of stronger residual ZF in D-T plasmas with larger $A_{i,eff}$ is weakened by non-trace impurities. This is because the presence of light or medium-mass impurities with finite concentration tends to weaken the influence from larger $\chi_{f_D D + f_T T, cl}$ for larger $A_{i, eff}$. In this region, the magnitude of impurity effects is related to $g_z \chi_{z, cl}/g_i \sim A_z f_c/(1 - Z f_c) \sim A_z f_c$ for $\tau_z = \tau_i$. Therefore, fully ionized impurities with heavier mass and higher $f_c$ have stronger influences on the reduction (enhancement) of the ratios $\frac{R_{ZF, z}^{0.1D+0.9T}}{R_{ZF, z}^{0.5D+0.5T}}$ and $\frac{R_{ZF, z}^{0.3D+0.7T}}{R_{ZF, z}^{0.5D+0.5T}}$ ( $\frac{R_{ZF, z}^{0.7D+0.3T}}{R_{ZF, z}^{0.5D+0.5T}}$ and $\frac{R_{ZF, z}^{0.9D+0.1T}}{R_{ZF, z}^{0.5D+0.5T}}$ ) as shown in figure 6 (a) and (b). While, in figure 7, the influences of trace W on the ratios are invisible because $g_z \chi_{z, cl}/g_i \sim A_z f_c$ for trace W is too small. In one word, trace W has very weak influences on the $A_{i, eff}$ dependence of residual ZF.

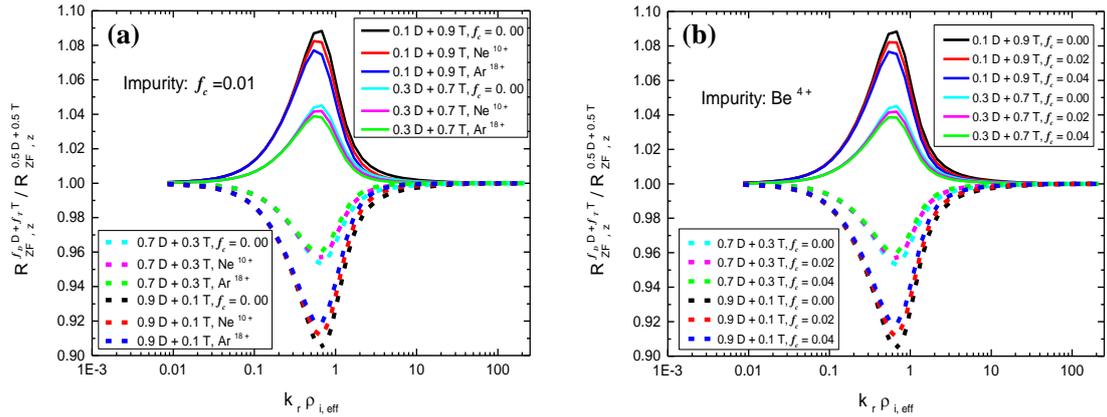

Figure 6. The ratios $\frac{R_{ZF, z}^{0.1D+0.9T}}{R_{ZF, z}^{0.5D+0.5T}}$, $\frac{R_{ZF, z}^{0.3D+0.7T}}{R_{ZF, z}^{0.5D+0.5T}}$, $\frac{R_{ZF, z}^{0.7D+0.3T}}{R_{ZF, z}^{0.5D+0.5T}}$ and $\frac{R_{ZF, z}^{0.9D+0.1T}}{R_{ZF, z}^{0.5D+0.5T}}$ as a function of $k_r \rho_{i, eff}$ for different fully ionized impurities with $f_c = 0.01$ in (a) and different $f_c$ with $Be^{4+}$ in (b). The lines with $f_c = 0.00$ correspond to the cases without impurities. Here, $\rho_{i, eff}$ is the effective ion gyroradius with $A_{i, eff} = 2.5$.

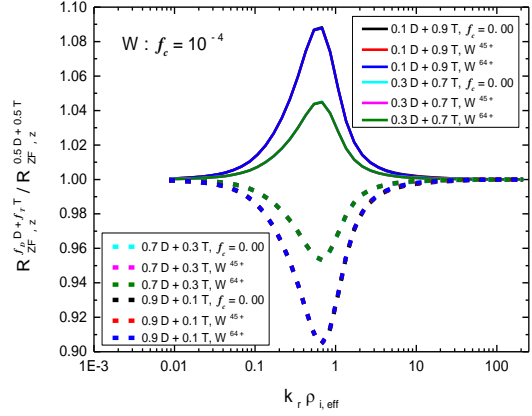

Figure 7. The ratios $\frac{R_{ZF,z}^{0.1D+0.9T}}{R_{ZF,z}^{0.5D+0.5T}}$, $\frac{R_{ZF,z}^{0.3D+0.7T}}{R_{ZF,z}^{0.5D+0.5T}}$, $\frac{R_{ZF,z}^{0.7D+0.3T}}{R_{ZF,z}^{0.5D+0.5T}}$ and $\frac{R_{ZF,z}^{0.9D+0.1T}}{R_{ZF,z}^{0.5D+0.5T}}$ as a function of $k_r\rho_{i,eff}$ for trace W with different Z. The lines with $f_c=0.00$ correspond to the cases without impurities. $\rho_{i,eff}$ is the effective ion gyroradius with $A_{i,eff}=2.5$.

Furthermore, we note that impurity effects on the $A_{i,eff}$ dependence of residual ZF might be relevant to the impurity effects on confinement of magnetic fusion plasmas. Gyrokinetic simulation in Ref. [24] showed that the hydrogenic isotope mass dependence of the linear growth rate of ITG is weakened by the presence of impurity. This is qualitatively consistent with our findings that non-trace impurities can weaken the $A_{i,eff}$ dependence of residual ZF.

## 4. Summary and discussions.

The influences of impurities on arbitrary wavelength residual ZF and on $A_{i,eff}$ dependence of residual ZF in D-T plasmas are investigated in this paper. The calculation of this work is simple, and the results are also very easily understood. However, we find significant influences of impurities on the levels of intermediate and short wavelength residual ZF, even for high-Z trace W, which are not addressed in previous works. The main results of this work are summarized in Table 1.

| Wavelength region | Impurity effects on residual ZF | | Impurity effects on $A_{i,eff}$ dependence of residual ZF |
| --- | --- | --- | --- |
| ITG-driven | Weak | | Weak |
| TEM-driven | $\rho_z > \rho_{i,eff}$: increase | Strengthened by increase of $\frac{\rho_z^2}{\rho_{i,eff}^2} \sim \frac{A_z}{A_{i,eff} Z^2} \frac{T_z}{T_i}$, $f_c$ | Weakened by non-trace impurities; Not affected by trace W. |
| ETG-driven | $\rho_z < \rho_{i,eff}$: decrease | Strengthened by increase of $\frac{\rho_{i,eff}^2}{\rho_z^2} \sim \frac{A_{i,eff} Z^2}{A_z} \frac{T_i}{T_z}$, $f_c$ | Weak |
| | $g_i + g_z > \tau_i$: increase | Strengthened by increase of $g_i + g_z - \tau_i = Zf_c(\tau_z Z - \tau_i)$ | |
| | $g_i + g_z < \tau_i$: decrease | Strengthened by increase of $\tau_i - (g_i + g_z) = Zf_c(\tau_i - \tau_z Z)$ | |

**Table 1.** Overview of the results about impurity effects on residual ZF and on $A_{i,eff}$ dependence of residual ZF in D-T plasmas.

Now, we compare our results with previous one, and discuss some possible implications for experiments and burning plasmas. Ref. [18] pointed out that the high-Z impurity does not affect the residual ZF. However, our results showed that it is only true for large scale RH residual ZF. We find that the relatively shorter scale residual ZF can be significantly affected by high-Z impurity even with trace concentration and by non-trace light or medium-mass impurities as well. It is also worth stressing that the influences of $He^{2+}$ from D-T reaction with different temperatures in different radial wavelength regions are complicated. Therefore, investigating multi-scale turbulence and ZF [38] with various impurities may be required for accurate estimation of impurity effects on confinement. Furthermore, we find that non-trace impurities can weaken the trend of stronger residual ZF level in D-T plasmas with higher $A_{i,eff}$, but trace W does not affect the trend. From experiment, the amplitude of long-range correlation (LRC) is increased during the transition from H to D plasmas in TEXTOR [39]. Therefore, it is also worth investigating the effects of impurities on the amplitude of LRC in different hydrogenic isotope plasmas by injecting medium-mass impurities. All the suggestions mentioned above can be also tested by gyrokinetic simulation by taking impurities into account.

Finally, it should be noted that an integrated study on complex interplay of the turbulence, ZF and impurities may be required for comprehensive understanding the overall effects of impurities on the performance of magnetic fusion plasmas. Our ongoing work focuses on ZF generation by collisionless trapped electron mode (CTEM) turbulence with impurities. Extension to study of ZF and zonal fields [1, 40, 41] in electromagnetic turbulence with impurities [42] might be also worthwhile.

We are grateful to P. H. Diamond for his suggestions. We also thank Y. Xu, Q. Yu, R. Singh, J. M. Kwon, H. Jhang and the participants in 6th Asia Pacific Transport Working Group (APTWG), Seoul, Korea 2016 for fruitful discussions. This work was supported by the NSFC Grant Nos. 11675059 and 11305071, the MOST of China under Contract No. 2013GB112002.